\documentclass{emulateapj}
\usepackage{graphicx}
\usepackage{amssymb,amsfonts,amsmath,amstext,amsgen,amsopn,amsxtra,indentfirst,times}
\usepackage{multirow,threeparttable,rotating}

\begin{document}

\title{Constraining the Atmospheric Composition of the Day-Night Terminators of HD 189733\lowercase{b} :  \\ Atmospheric Retrieval with Aerosols}
\author{Jae-Min Lee\altaffilmark{1,2}}
\author{Patrick G. J. Irwin\altaffilmark{3}}
\author{Leigh N. Fletcher\altaffilmark{3}}
\author{Kevin Heng\altaffilmark{2}}
\author{Joanna K. Barstow\altaffilmark{3,4}}
\altaffiltext{1}{University of Z\"{u}rich, Institute for Computational Science, Winterthurerstrasse 190, CH-8057, Z\"{u}rich, Switzerland. Email: lee@physik.uzh.ch}
\altaffiltext{2}{University of Bern, Center for Space and Habitability, Sidlerstrasse 5, CH-3012, Bern, Switzerland.}
\altaffiltext{3}{University of Oxford, Department of Atmospheric, Oceanic and Planetary Physics, OX1 3PU, Oxford, UK.}
\altaffiltext{4}{University of Oxford, Department of Astrophysics, OX1 3RH, Oxford, UK}

\begin{abstract}
A number of observations have shown that Rayleigh scattering by aerosols dominates the transmission spectrum of HD 189733b at wavelengths shortward of 1 $\mu$m. In this study, we retrieve a range of aerosol distributions consistent with transmission spectroscopy between 0.3--24 $\mu$m that were recently re-analyzed by \citet{pon13}.  To constrain the particle size and the optical depth of the aerosol layer, we investigate the degeneracies between aerosol composition, temperature, planetary radius, and molecular abundances that prevent unique solutions for transit spectroscopy. Assuming that the aerosol is composed of MgSiO$_3$, we suggest that a vertically uniform aerosol layer over all pressures with a monodisperse particle size smaller than about 0.1 $\mu$m and an optical depth in the range 0.002--0.02 at 1 $\mu$m provides statistically meaningful solutions for the day/night terminator regions of HD 189733b. Generally, we find that a uniform aerosol layer provide adequate fits to the data if the optical depth is less than 0.1 and the particle size is smaller than 0.1 $\mu$m, irrespective of the atmospheric temperature, planetary radius, aerosol composition, and gaseous molecules. Strong constraints on the aerosol properties are provided by spectra at wavelengths shortward of 1 $\mu$m as well as longward of 8 $\mu$m, if the aerosol material has absorption features in this region.  We show that these are the optimal wavelengths for quantifying the effects of aerosols, which may guide the design of future space observations. The present investigation indicates that the current data offer sufficient information to constrain some of the aerosol properties of HD189733b, but the chemistry in the terminator regions remains uncertain.
\end{abstract}


\section{Introduction}
\label{s1}

The dawn and dusk terminators of HD 189733b represent some of the best-studied atmospheric regions among the known transiting exoplanets, with the spectral coverage ranging from the ultraviolet (UV) to the mid-infrared (MIR).  However, there remain discrepancies between spectra measured by the same instruments  (specifically, the planet-star radius ratio as denoted by $R_{p}/R_{s}$), including: (i) disagreement in $Spitzer$/IRAC measurements \citep{bea08,des09,ago09,knu07}, thought to be due to different data reduction processes applied to the same transit measurements \citep{for10,sha11}; and (ii) disagreement in $HST$/NICMOS observations due to different approaches to transit curve decorrelation \citep{swa08,sin11,gib11,gib12a,gib12b}. The discrepancies between different groups stem partly from difficulties in unequivocally correcting for instrumental effects. 

At wavelengths shorter than 1 $\mu$m, the transmission spectrum decreases monotonically from the UV to the near-infrared (NIR), as measured by $HST$/ACS (0.55--1.05 $\mu$m) \citep{pon08} and $HST$/STIS (0.29--0.57 $\mu$m) \citep{sin11}.  This sloped spectrum is consistent with the presence of particles in the Rayleigh scattering regime and suggests the presence of aerosols in the terminator regions. \citet{lec08} suggested that a thick and vertically extended aerosol layer consisting of silicate grains such as enstatite (MgSiO$_3$) or forsterite (Mg$_2$SiO$_4$) is able to reproduce the Rayleigh slope. By applying stellar spot corrections for all available measurements between 0.3--24 $\mu$m, \citet{pon13} suggested that the opacity from an optically thick aerosol layer governs the entire transmission spectrum and that the poor quality of the present data cannot be used to identify molecular signatures in the terminators.  Given this uncertainty in the data, and the degeneracies inherent in their reduction and analysis, the atmospheric properties for the terminators differ from study to study.

In this study we revisit the atmosphere of the day/night terminators of HD 189733b, considering the transmission spectrum reassessed by \citet{pon13}, where the maximum information content from the previous transit data were extracted. It clearly reveals a precipitous slope in the $<$1 $\mu$m spectrum and a flatter spectrum in the IR. In particular, the spectrum near 1 $\mu$m is particularly important in determining which opacity source is more plausible in the terminators in the IR, i.e., optically thick aerosols washing out all other molecular signatures \citep{des09,sin11,gib12a,pon13}; or a combination of aerosols and molecules \citep{tin07b,bea08,swa08}. The main objective of this study is therefore to assess the family of plausible aerosol models that are consistent with the measurements currently available by exploring degeneracies between aerosol and other atmospheric parameters, i.e., temperature, planetary radius and molecular composition. We also aim to investigate the influence of different types of aerosols on the transmission spectrum between 0.3--24 $\mu$m, demonstrating the importance of measuring a wide spectral range to break the degeneracies between competing atmospheric parameters.

Atmospheric properties were derived using the NEMESIS spectral retrieval tool \citep{irw08}, seeking the best-fits across a pre-defined parameterized space of aerosol properties: nadir (or radial) optical depth ($\tau$) and particle radius ($a$). The vertical locations and extent of the aerosol layers are poorly characterised with existing measurements, so we choose to define a simple and phenomenological aerosol model following \citet{lee13}, where aerosols are uniformly distributed throughout the pressures of the model atmosphere (10$^{-7}$--10 bar) (UC, the number of particles per cubic centimeter is vertically uniform).  This allows us to reduce the number of free parameters during the retrieval process.  More sophisticated aerosol models are commonly used for the study of sub-stellar objects \citep[e.g.][]{ack01,hel08,mad11,bar11,mar12}, but we find that the existing data quality and low signal-to-noise does not warrant the use of more complex aerosol models at this time. In this study, we explore broad ranges for two aerosol parameters, i.e., $\tau$ and $a$ for various atmospheric conditions so that we are able to characterise the degeneracies between the aerosol properties and other parameters, e.g., thermal structure, planetary radius, aerosol composition, gaseous compositions.

\section{Modelling}
\label{s2}

\subsection{Retrieval}
NEMESIS finds the best-fitting atmospheric state (composition and aerosol properties) that reproduces the observed transit spectrum iteratively, where the next step of retrieval is analytically calculated using the technique of optimal estimation \citep{rod00,irw08}. This spectral inversion software was modified to permit analyses of exoplanet spectroscopy and to seek the range of plausible solutions for exoplanetary atmospheres with realistic uncertainties \citep{lee12,bar13,bar14,lee13}. To achieve the aerosol information that can be derived from the given data, we grid $\tau$ and $a$ over a wide range of values (see Section~\ref{s22}). We consider a range of molecules and atoms (H$_2$O, CO, CO$_2$, CH$_4$, Na and K) that have signatures in the visible and IR. For the sources of opacity, we used the line lists of HITEMP for H$_2$O, CO, and CO$_2$\citep{hitemp10}, STDS for CH$_4$\citep{stds} as explained in \citet{lee12}, and VALD for Na and K\citep{vald}. These opacity sources are the only fitting variables that are being retrieved at each grid point of the aerosol models. Assuming that molecular and atomic abundances are well-mixed with altitude, we define a single scale factor for these profiles as fitting variables in the retrieval. H$_2$ and He are considered to have a fixed abundance ratio, $X_{He}/X_{H_2}=$ 0.097, i.e., assuming that they have solar abundance, during retrieval. For every iteration of the retrieval the summation of the volume mixing ratio of all species must be unity, i.e., $\sum X_{molecules}$=1, to obtain physically meaningful solutions, where the condition of $1-\sum{X_{H_2O, CO_2, CO, CH_4, Na, K}}=\sum{X_{H_2,He}}$ is always satisfied.

\begin{figure}
 \centering
 \includegraphics[width=\columnwidth]{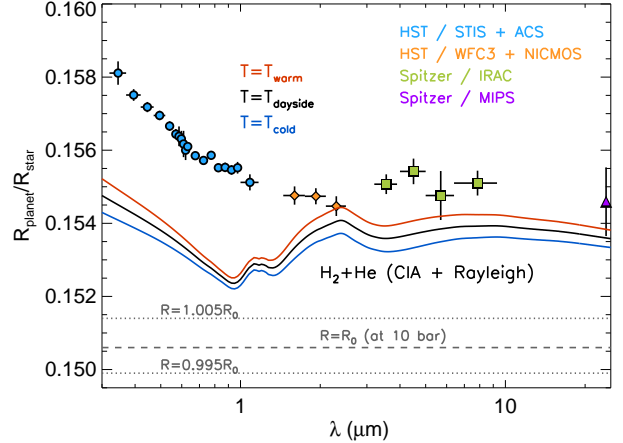}
 \caption{The transmission spectrum of HD 189733b, reanalysed by \citet{pon13} using data taken from the $HST$ and $Spitzer$ space telescopes.  A data point at 589.5 nm is not considered here due to its resolution, $\Delta\lambda$=2.2 nm, which is narrower than the 5 nm spectral resolution of our spectral models. The bottom level of the model atmosphere is shown in the dashed line and equivalent to the radius of the planet at 10 bar, with which we can find the best-fitting solution for the IR measurements. See main text for more details. The upper and lower limits of $R_0$ are displayed as dotted lines. At $R=R_0$, the continuum due to collisional induced absorption and Rayleigh scattering by H$_2$ and He are shown with $T_{dayside}$ (black), $T_{dayside}+200$K ($=T_{warm}$) (red) and $T_{dayside}-200$K ($=T_{cold}$) (blue) (see the context for more detail). A Rayleigh slope by H$_2$ and He may be able to fit the $HST$/STIS and ACS data points if the planetary radius has a large uncertainty. {\it However, we do not consider this scenario in the present study.} }
\label{f1}
\end{figure}

Since transmission spectra have a low sensitivity to the vertical temperature profile \citep{tin07a,ben12,bar13}, we have insufficient constraints for an independent retrieval of the vertical structure of temperature. However, the atmospheric scale height, which mainly determines the depth of the transmission spectrum, is a function of temperature, implying the necessity of the exploration of the temperature degeneracy with aerosol parameters and chemical composition. Hence, we assume that the day/night terminators have the same P-T profile as the dayside atmosphere \citep{lee12}. Specifically, we test the robustness of our retrieved composition and aerosol properties by repeating the analysis with the temperatures shifted by $\pm$200~K either side of the best-fitting P-T profile of the dayside hemisphere ($T=T_{dayside}$) (Figure~\ref{f1}). The details are discussed in Section~\ref{s321}. 

In addition to temperature degeneracy, we also consider the planetary radius ($R_p$) at the bottom level of the model atmosphere: $R_p=R_{10 bar}$. This is an additional free parameter adding considerable uncertainty to atmospheric parameters during retrieval. This is because a change in the radius of planet, which defines the transit depth of the bottom of the model atmosphere, results in a change in compositional amount of molecules and aerosols \citep[e.g.][]{bar13,ben13}. In other words, retrievals with a smaller $R_p$ favor opacity increased due to enhanced molecular abundances or aerosol amount, whereas a larger $R_p$ requires decreased opacity due to reduced amount of composition to fit the transit measurements. 

For the dayside emission spectrum of HD 189733b, \cite{lee12} used $R_p=R_0$ (=$1.138R_J$), based on that measured in the optical \citep{tor08}, as a model input, where $R_p/R_s=$0.15463$\pm$0.00022. In the present analysis, we perform an additional retrieval to determine $R_0$ using the fixed best-fit temperature profile and molecules from the dayside atmosphere. We fitted the IR measurements including NICMOS, IRAC and MIPS channels and obtained the best-fit planet-to-star ratios ($R_{0}/R_s$) of 0.1506 for $T=T_{dayside}$, which is smaller than the value in \citet{tor08}. Furthermore, we assigned the upper and lower limits to $R_0$ so that $1.005R_{0}/R_s=0.1514$ and $0.995R_{0}/R_s=0.1499$, an uncertainty envelope that is seven times larger than suggested by\citep{tor08}. We investigate the sensitivity of the retrieved composition and aerosol properties by varying the P-T profile and planetary radius within these conservative uncertainty ranges. For more details, we will show our results on the degeneracy due to the planetary radius in Section~\ref{s321}.

\begin{figure}
\centering
\includegraphics[width=\columnwidth]{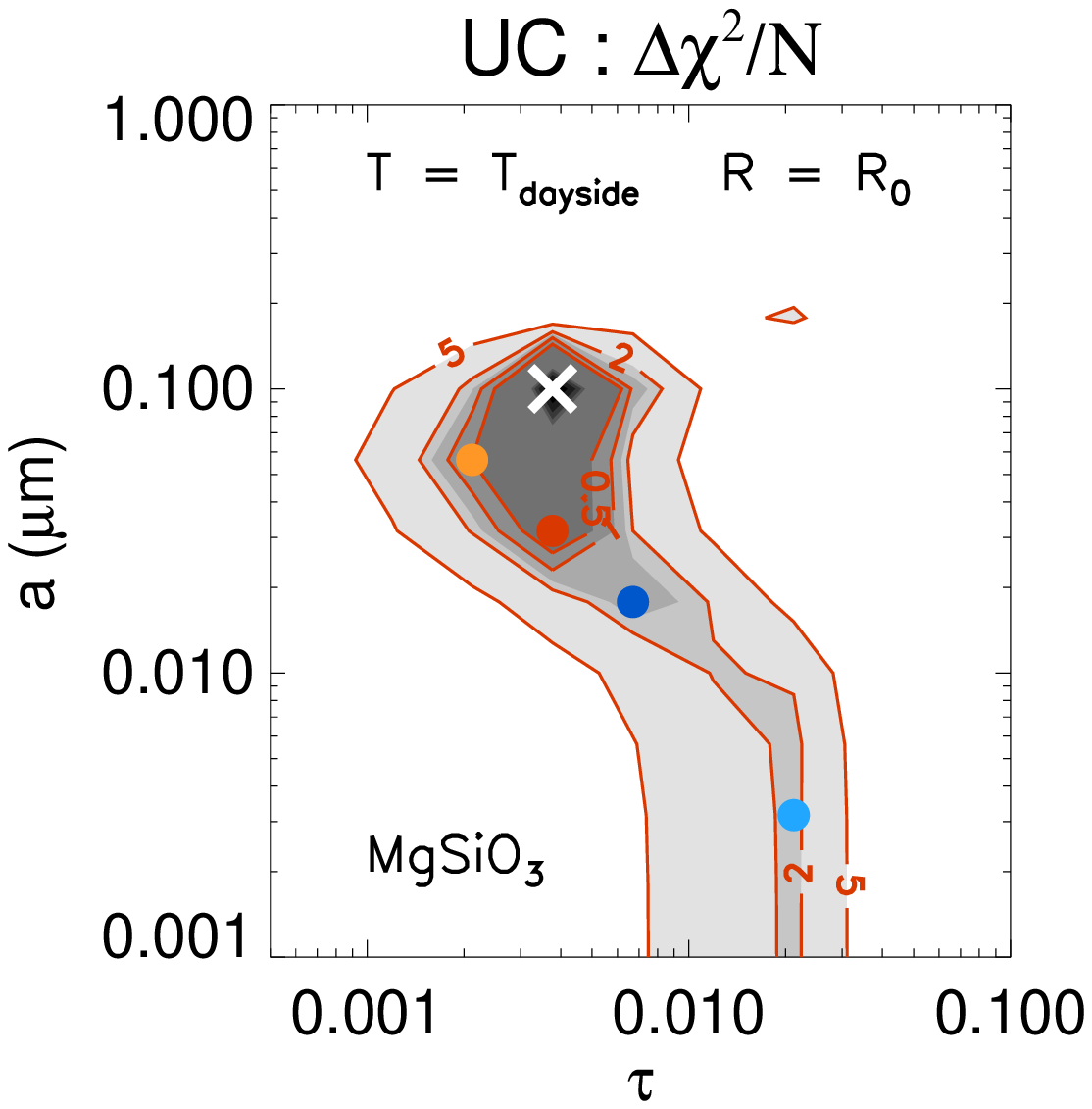}
\includegraphics[width=\columnwidth]{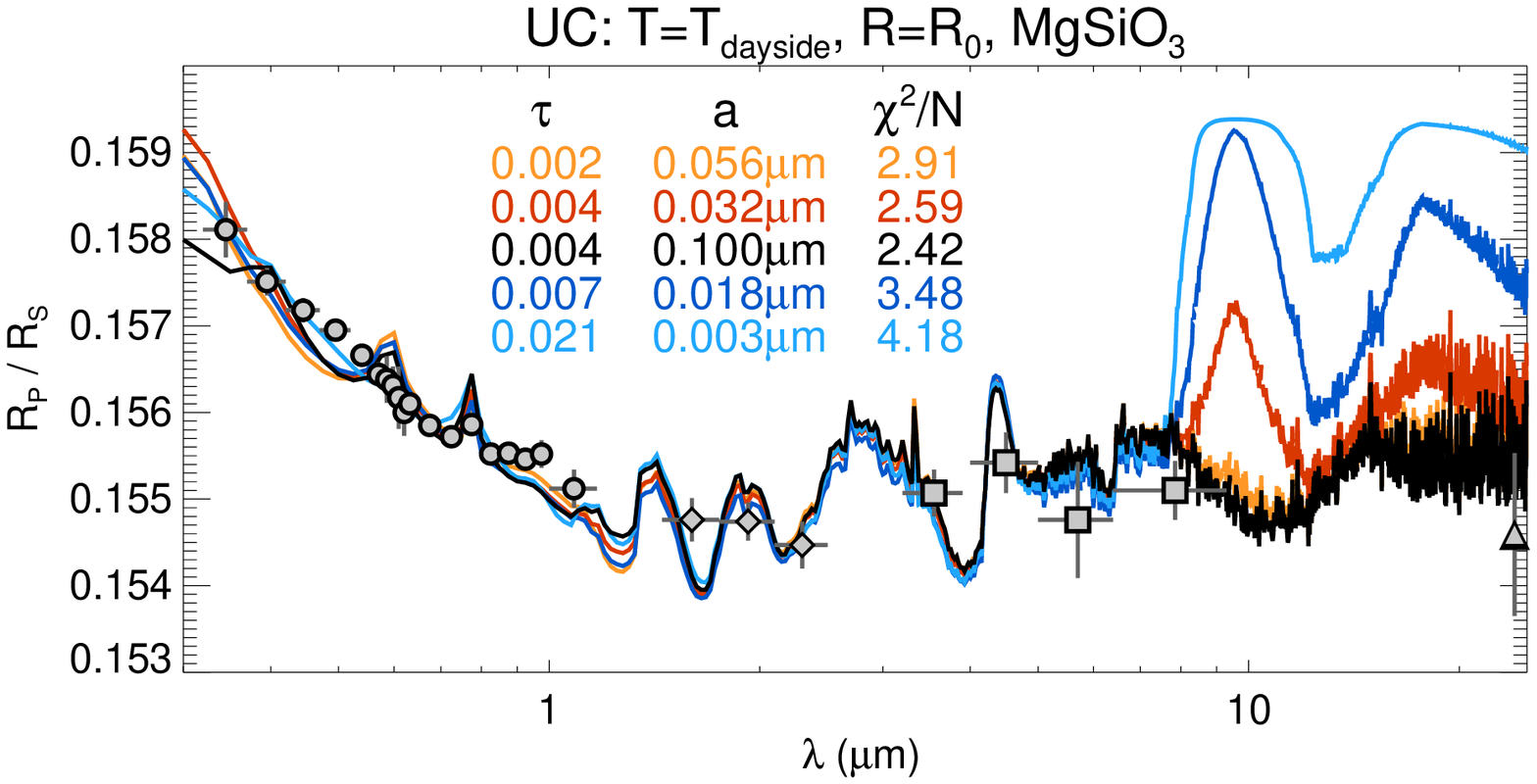}
\caption{In the top panel, the goodness-of-fit, $\Delta\chi^2/$N, contours for the uniform aerosol models with single particle sizes are shown, where $T=T_{dayside}$ and $R_p=R_0$. The best fit is marked with white cross -- corresponding spectrum is shown in black in the bottom panel. The contour plot demonstrates a constrained upper boundary of particle size ($a\sim0.1\mu$m) despite of the degeneracy with the aerosol optical depth ($\tau$). The bottom panel shows differences between the spectra, where each color corresponds to the points in the contour above. Discrepancies appear mostly longward of 8 $\mu$m, where the strong extinction by silicate is shown. At $>$8 $\mu$m, silicate is responsible for the spectral morphology and degrades the fitting quality for a region where aerosol has small $a$ and thick $\tau$ in the top panel.}
\label{f2}
\end{figure}

\subsection{Aerosol}
\label{s22}
The nadir optical thickness due to the extinction by an aerosol layer is given by,
\begin{align}
\tau = \rho_n~\Delta z~\pi a^2~Q_{ext}, \nonumber
\end{align}
where $\rho_n$ is the number density in cm$^{-3}$, $\Delta z$ is the column height and $a$ is the radius of the spherical particle.  $Q_{ext}$ is the extinction efficiency of the grain, a sum of the absorption efficiency ($Q_{abs}$) and the scattering efficiency ($Q_{sca}$). Since we consider only the first approximation of a single-scattering calculation, we assume that a photon that is either absorbed or scattered by a particle is not returned to the line of sight. We consider that $\rho_n$ is vertically uniform in between the pre-defined pressures of 10$^{-7}$ bar and 10 bar, which corresponds to about 15 atmospheric scale heights for $T=$1500~K and a mean molecular weight of 2.3. As in \citet{lee13}, these pressure levels are defined to be the top and bottom of the model atmosphere for the vertically uniform aerosol model (UC) and the pressure levels covers the range of aerosol heights \citep[e.g. $\lesssim$10$^{-5}$ bar,][]{sin11}. We choose not to vary these upper and lower boundary pressures in this study because the main objective of this study is to explore the general influence of aerosols on the retrieval of the temperature, planetary radius, and molecular degeneracies consistent with the spectrum, irrespective of aerosol layer geometry.

Formally, the aerosol properties should be part of the retrieval.  However, as a first approach, we fix $\tau$ and $a$ and fit the spectrum by varying the gaseous abundances for each retrieval and perform a suite of retrievals over a broad range of $a = 0.001$--10 $\mu$m and $\tau=0.0001$--10. This approach allows us to explore the parameter space associated with the aerosol properties and to understand the inherent degeneracies. 

In the next section, we will derive the best-fit scenario with an aerosol layer consisting of MgSiO$_3$ and explore how the aerosol properties are correlated with temperature, planetary radius, other candidate materials for the aerosols, and gaseous composition.

\section{Results}
\label{s3}

\subsection{Best-fit model}
\label{s31}

We perform the retrieval with the full transmission spectrum of HD 189733b between 0.3--24 $\mu$m, using the ranges of $\tau$ and $a$ described in section~\ref{s22}. In this first study, the aerosol is modelled as enstatite (MgSiO$_3$) grains \citep{sco96}, which is a potential candidate for aerosols in sub-stellar atmospheres \citep[e.g.][]{bur99,lod99,hel06,lec08,mor12,pon13}. The temperature and radius ratio are fixed to $T_{dayside}$ and $R_0$ for this case. The fitted spectra are assessed using the goodness-of-fit, i.e. weighted least mean square error, $\chi^2/$N, where N=27 is the number of measurements. The fitting parameters include H$_2$O, carbon chemicals (CO, CO$_2$ and CH$_4$), and alkalis (Na and K). Since opacities for carbon-based molecules at high temperatures are either unavailable or weak for wavelengths shorter than 1 $\mu$m, H$_2$O, Na, K and extinction by MgSiO$_3$ are predominant in the $HST$/STIS and ACS bandpasses. Figure 2 illustrates the contours of $\Delta\chi^2$/N for a range of $\tau$ and $a$. For clarity, we only illustrate reduced ranges of aerosol parameter space throughout the study due to poor fits for all $\tau$ and $a$ larger than 0.1 and 1 $\mu$m, respectively. The aerosol models having a particle radius smaller than $\sim$0.1 $\mu$m are able to fit the data for $\Delta\chi^2/$N ($=\chi^2/$N$-\chi^2_{minimum}/$N) $<$2, provided that $\tau$ lies between 0.002--0.02. The retrievals effectively rule out any scenario involving particles larger than $\sim$0.1 $\mu$m in radius, with $\tau$ outside of the 0.002--0.02 range. 

\begin{figure}
\includegraphics[width=\columnwidth]{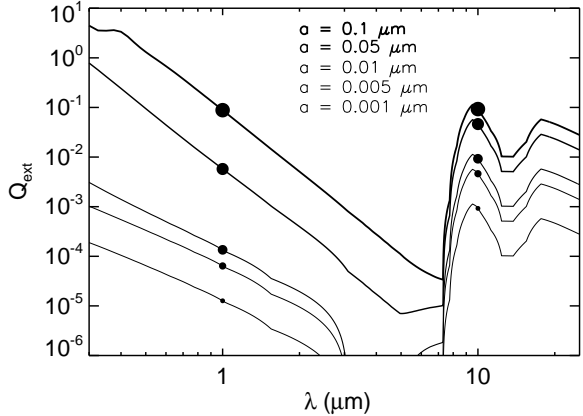}
\caption{Extinction efficiency of MgSiO$_3$ as a function of wavelength. An increase in $Q_{ext}$ at 1 $\mu$m due to an increase in $a$ is more sensitive than an increase at 10 $\mu$m. Given the same $\tau$ at 1 $\mu$m, the smaller $a$ is, the greater $\tau$ is expected at wavelengths longer than 8 $\mu$m, where an upturn in $Q_{ext}$ is shown.} 
\label{f3}
\end{figure}

In the bottom panel in Figure~\ref{f2}, it is found that different combinations of $\tau$ and $a$ are indistinguishable up to wavelengths of 8 $\mu$m for all fitted spectra, the overall shape of which shows a decreasing absorption slope due to an extinction by aerosols combined with a feature of Na, K and H$_2$O at wavelengths shorter than 1 $\mu$m and molecular absorptions by H$_2$O, CO, CO$_2$ and CH$_4$ at wavelengths between 1--8 $\mu$m. Differences between the radius and optical depth combinations can only be seen in spectra at longer wavelengths, where a strong absorption by MgSiO$_3$ is expected. This is the reason for an anti-correlation between $\tau$ and $a$ as shown in the contour plot, i.e., the larger the size of the particles, the smaller the opacity required to fit the spectrum beyond 8 $\mu$m. In Figure~\ref{f3}, we find that, for a fixed $Q_{ext}$ (or $\tau$) at 1 $\mu$m, an aerosol layer with $a\sim$ 0.1 $\mu$m shows a smaller $\tau$ at $>$8 $\mu$m than that with small $a<$0.1 $\mu$m due to the intrinsic characteristic of $Q_{ext}$ for MgSiO$_3$. As a result, the measurements of $Spitzer$/MIPS and IRAC at 8 $\mu$m place strong constraints that move $\tau$ to be decreased and $a$ to be increased by an order of magnitude in the contour plot. This shows that the spectrum at wavelengths $>$8 $\mu$m contains a large amount of information about the aerosols as seen at wavelengths $<$1 $\mu$m. This is, again, due to the wavelength-dependent nature of the $Q_{ext}$ of MgSiO$_3$. The dependence of $Q_{ext}$ on the fitting quality and aerosol properties will be discussed in Section~\ref{s322}.

An aerosol layer composed of small particles ($a/\lambda<$ 1) produces a negligible opacity between 1--8 $\mu$m, where molecules, primarily H$_2$O, are able to account for the $HST$/NICMOS and $Spitzer$/IRAC spectrum.  This implies that any conclusions for aerosols made in this study are unaffected by the diverse interpretations of the $HST$/NICMOS data by different studies \citep{swa08,sin11,gib11,gib12a,gib12b}.  It is also possible that the large errors on measurements in the 1--8 $\mu$m region permit solutions where the flat spectrum could be produced by a thick aerosol layer with large particles ($a/\lambda \geq$1), for which $Q_{ext}$ is nearly flat across wavelengths.  In this sense, \citet{sin11,pon13} suggested that the whole transmission spectrum could be reproduced if small grains at high altitude are responsible for the spectrum at wavelengths $<$1 $\mu$m, and deeper, larger grains ($a\sim$ 1$\mu$m) underlying the lower aerosol height forming the flat spectrum across the IR.  Although more complex solutions with multiple particle sizes may be considered, such complexity is not required to fit the data.

For the best-fit scenario for the full spectral range, we infer $a=$ 0.1 $\mu$m and $\tau=$ 0.004\footnote{We here quote nadir optical depth, but the slant path optical depth as probed
by transmission spectroscopy is much larger.}. Molecular abundances in the terminator regions are comparable with theoretical predictions of chemistry--the H$_2$O and CO abundances ($X_{H_2O},\ \ X_{CO}\approx$10$^{-4}$) are consistent across the globe ($X_{H_2O}\approx10^{-4}$ for both hemispheres), and a high CO$_2$ abundance and a low CH$_4$ abundance ($X_{CO_2},\ \ X_{CH_4}\approx$10$^{-6}$) are found compared to a disequilibrium chemistry model ($X_{CO_2}\approx10^{-7}$ and  $X_{CH_4}\approx10^{-5}$) \citep[e.g.][]{mos11}, resulting in a low C/O ratio compared to the dayside atmosphere, which is close to the solar value of about 0.5--0.6 \citep{mad09,lee12,lin13}. However, the best-fitting solutions for molecules listed here have a broad uncertainty due to degeneracies with temperature, planetary radius, and aerosol composition. Therefore, uncertainties in these best-fitting values will be assessed in subsequent sections.

\subsection{Assessing uncertainties in aerosol properties}

\begin{figure*}
\centering
\includegraphics[width=\columnwidth]{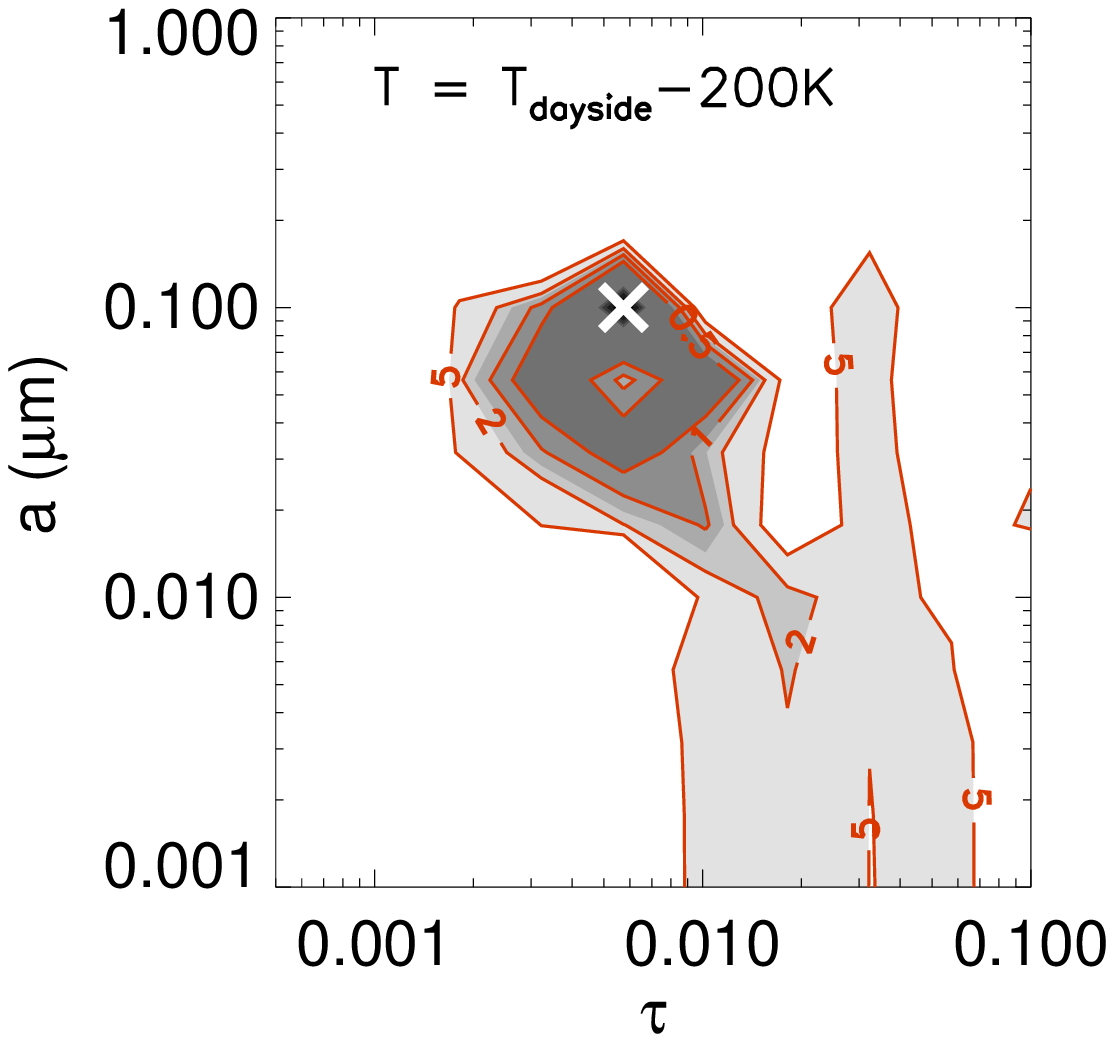}
\includegraphics[width=\columnwidth]{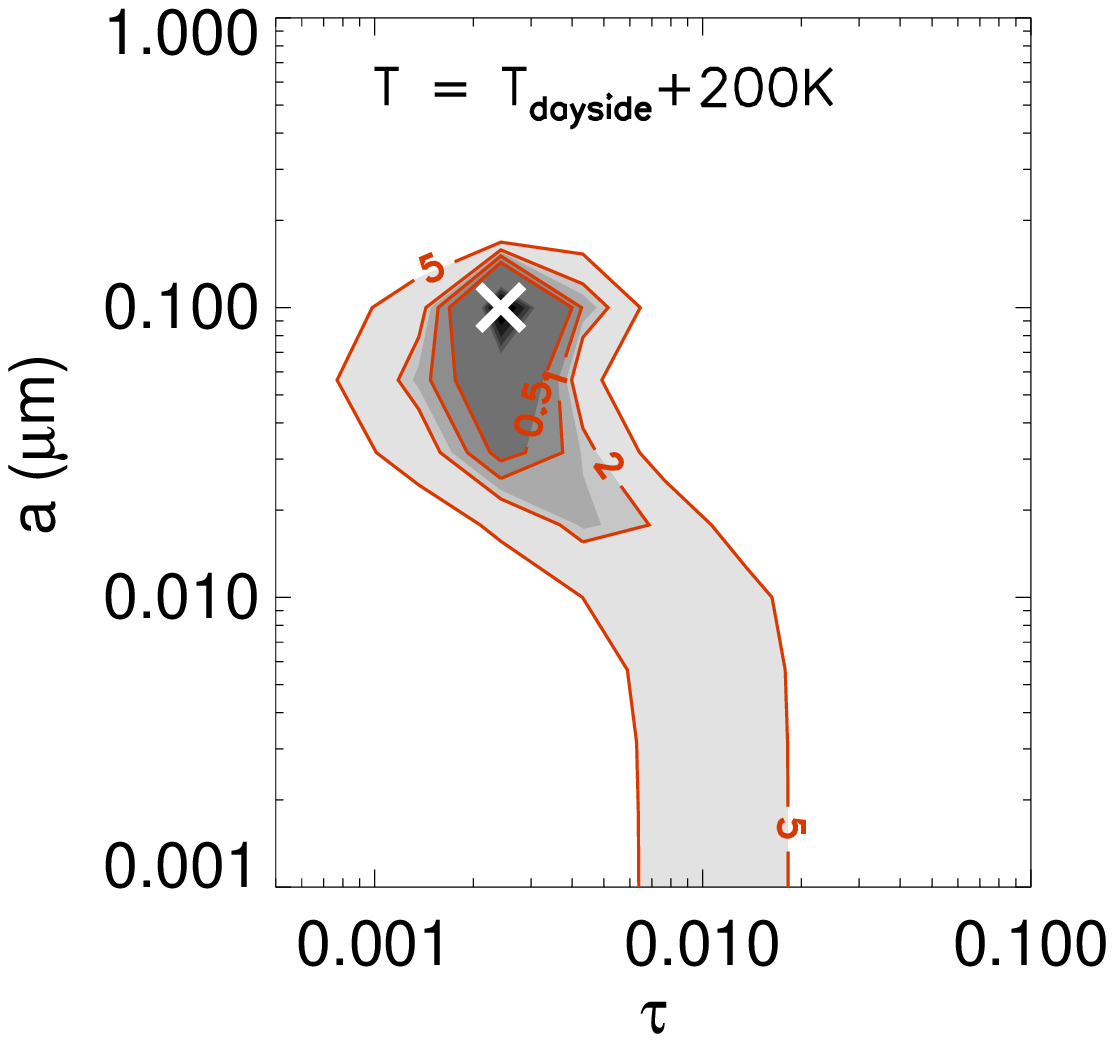}
\caption{Given $R=R_{0}$, the goodness-of-fit, $\Delta\chi^2/$N, contours in aerosol $\tau$ and $a$ space for the vertical temperature profiles of $T_{cold}$ and $T_{warm}$. It is found that a thinner aerosol $\tau$ is needed for the warmer case. The upper limit of $a$ ($\sim$0.1 $\mu$m) for $T=T_{cold}$ and $T=T_{warm}$ is consistent with $T=T_{dayside}$.}
\label{f4}
\end{figure*}

\begin{figure*}
\centering
\includegraphics[width=\columnwidth]{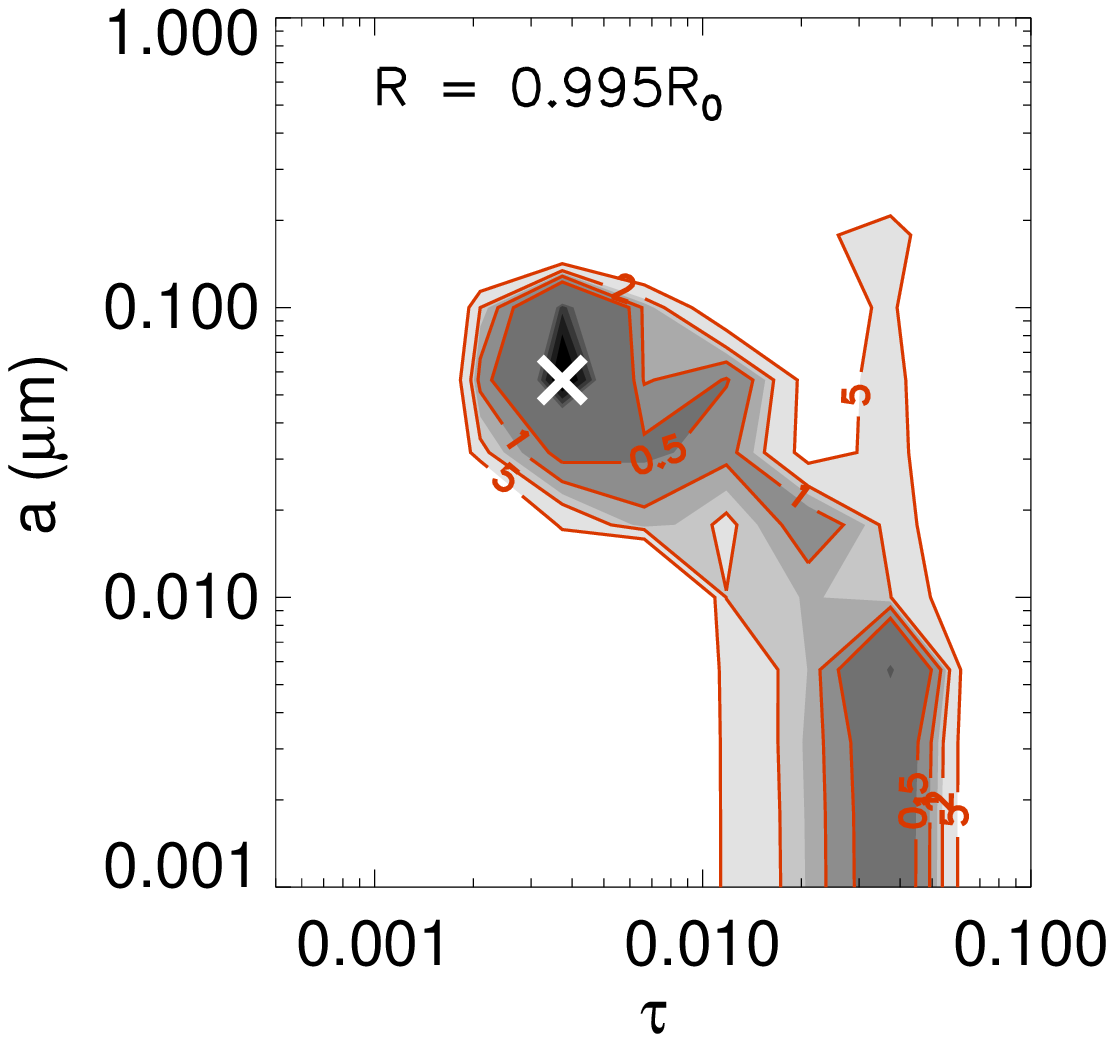}
\includegraphics[width=\columnwidth]{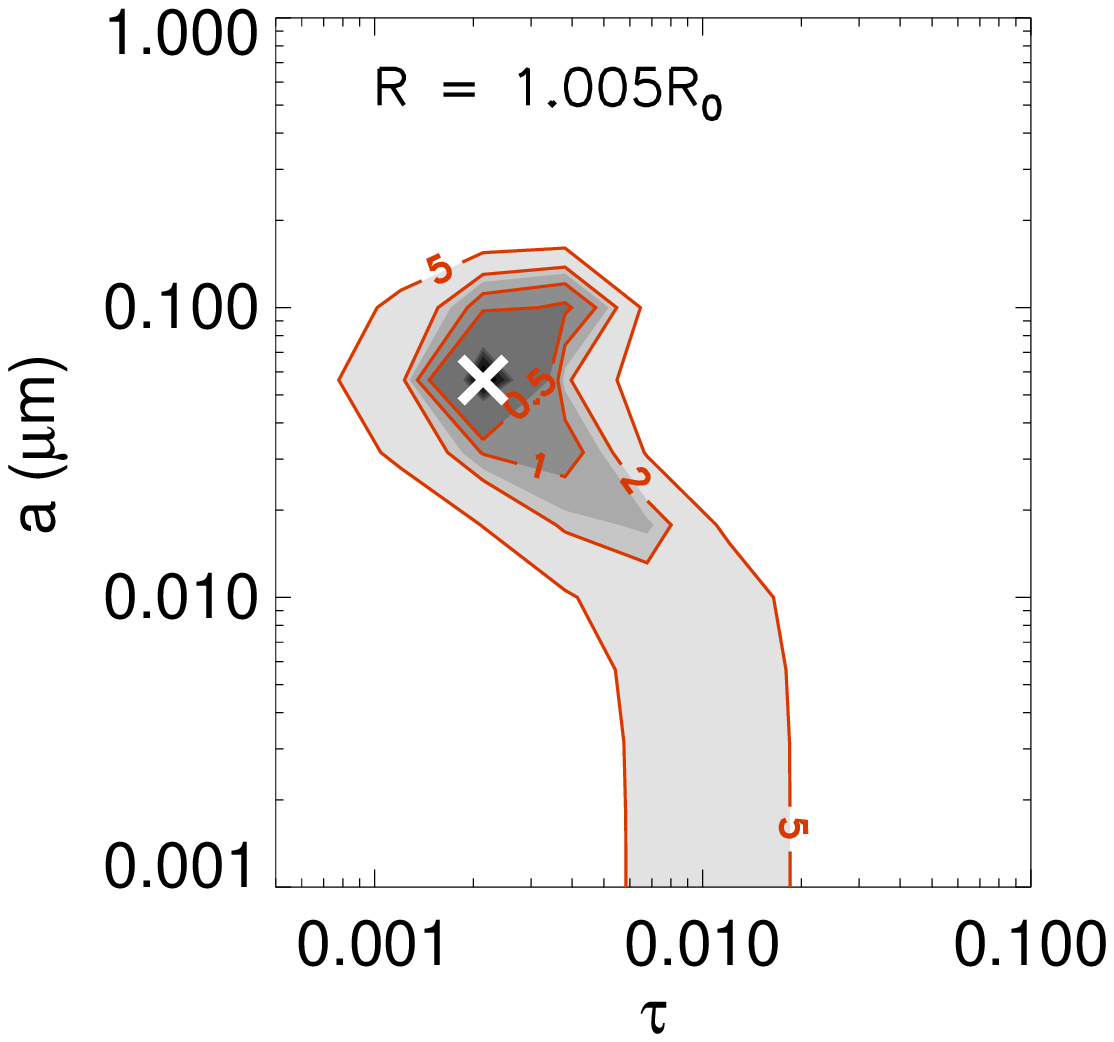}
\caption{Given $T=T_{dayside}$, the goodness-of-fit, $\Delta\chi^2/$N, contours in aerosol $\tau$ and $a$ space for radii moved up and down by 0.5\% from $R_{0}$. It is found that a thinner aerosol $\tau$ is needed for the larger $R_p$. Again, the upper limit of $a$ ($\sim$0.1 $\mu$m) for $R_p=0.995R_0$ and $R_p=1.005R_0$ is still consistent with $R_p=R_0$. }
\label{f5}
\end{figure*}

\begin{figure*}
\centering
\includegraphics[width=1.5\columnwidth]{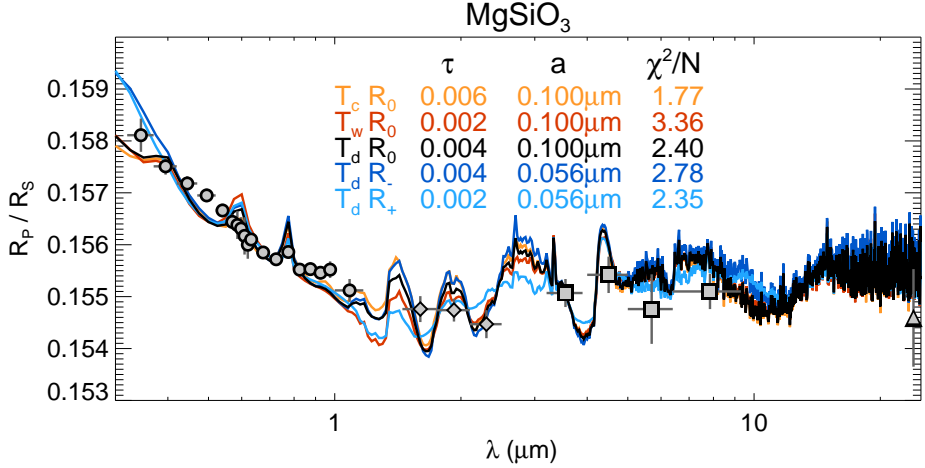}
\caption{The best-fit spectra retrieved from the five retrieval cases with $T=T_{cold}$, $R_p=R_0$ (orange), $T=T_{warm}$, $R_p=R_0$ (red), $T=T_{dayside}$, $R_p=R_0$ (black), $T=T_{dayside}$, $R_p=0.995R_0$ (blue) and $T=T_{dayside}$, $R_p=1.005R_0$ (cyan). The largest deviation between the spectra is shown at wavelengths 1--3.5 $\mu$m, where the H$_2$O features appear. } 
\label{f6}
\end{figure*}

\begin{figure*}
\centering
\includegraphics[width=1.5\columnwidth]{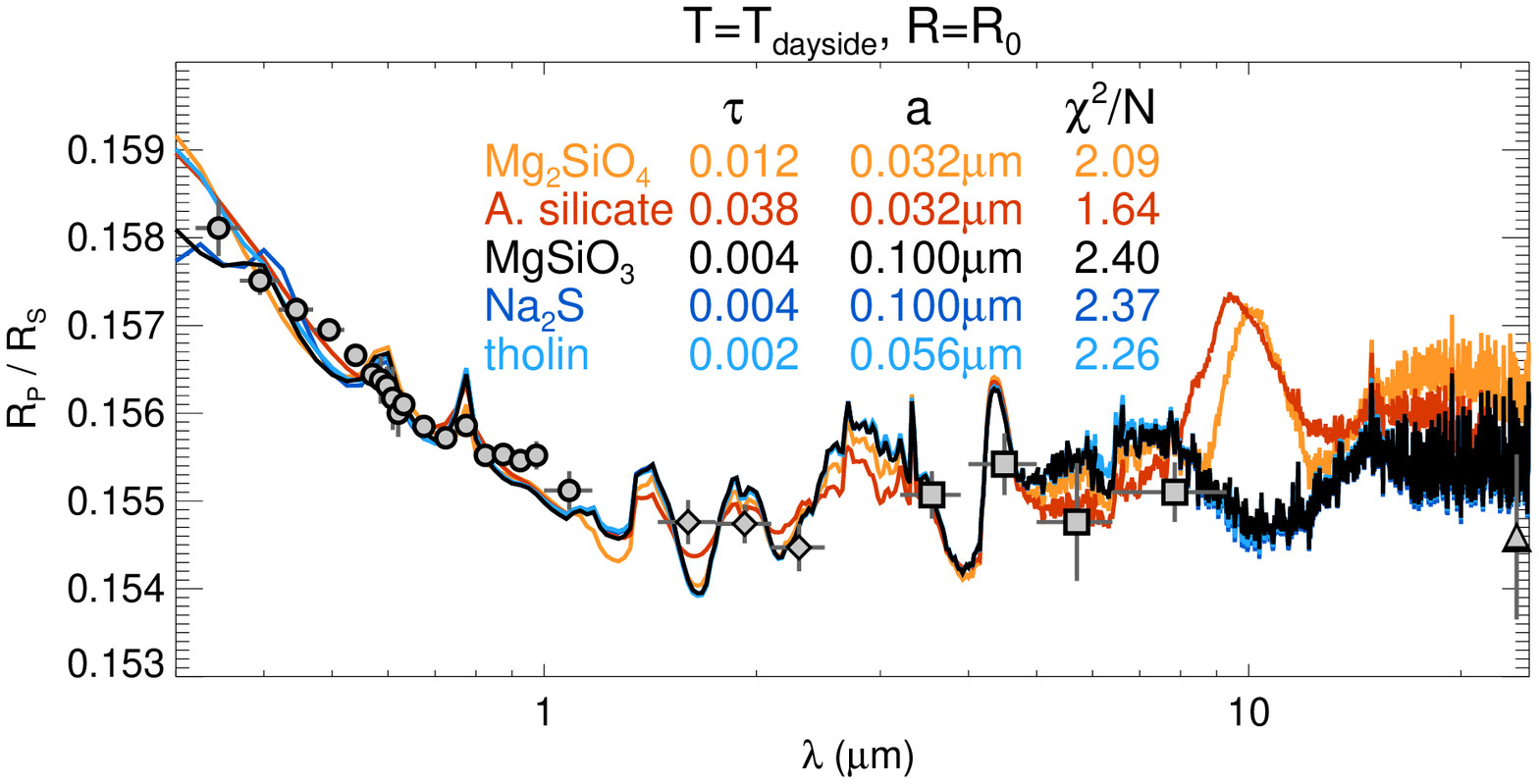}
\caption{The best-fit spectra retrieved from five retrieval cases with aerosols consisting of Mg$_2$SiO$_4$ (orange), astronomical silicate (red), MgSiO$_3$ (black), Na$_2$S (blue) and tholin (cyan). } \label{f7}
\end{figure*}

\subsubsection{Temperature and Radius}
\label{s321}
During retrieval, temperature and planetary radius change opacity, which defines the absorption feature at each wavelength. Given the fundamental degeneracy between how these parameters affect the spectrum, however, the retrieval of these parameters is not independently constrainable \citep{bar13,ben13}. As a result, a change in opacity due to temperature and planetary radius alters the aerosol properties significantly. In this section, we will therefore investigate how the effects of these degeneracy sources change the aerosol solution.

Firstly, to test the sensitivity of the aerosol properties to temperature, we repeat the aerosol retrieval analysis for colder and warmer thermal structures, i.e., $T_{cold}$ and $T_{warm}$, for which the temperature of the entire profile is shifted by 200K from the dayside temperature profile ($T_{dayside}$) as introduced in Section~\ref{s22}. The planetary radius is fixed to $R_p=R_0$. Setting up the same retrieval parameters listed in Section~\ref{s22}, suites of $\tau$ and $a$ with $T_{cold}$ and $T_{warm}$ are explored. In Figure~\ref{f4}, it is shown that the colder temperature profile increases $\tau$ compared to the $T_{dayside}$ case whereas the warmer temperature shows a thinner $\tau$ than the $T_{dayside}$ case. This is because warmer temperature (1) increases the atmospheric scale height and (2) enhances the continuum absorption from the H$_2$--He CIA and Rayleigh scattering across wavelengths  (see Figure~\ref{f1}). An enhanced continuum absorption by $T_{warm}$, in turn, results in reduced molecular abundances (particularly H$_2$O) compared to the $T=T_{dayside}$ case that is necessary to fit the $HST$/NICMOS and $Spitzer$/IRAC fluxes. However, this yields a poor fit to the $HST$/ACS channels in the 0.8--1.1 $\mu$m (spectrum in red color in Figure~\ref{f6}), resulting in the best fitting quality for $T=T_{warm}$ worsened over $T=T_{dayside}$. 

For $T=T_{cold}$,  the continuum absorption by H$_2$ and He is reduced and the molecular abundances are enhanced to fit the measurements in the IR. Since $T=T_{cold}$ causes the scale height of atmosphere to be reduced, which reduces the scale of the spectral features, the best-fit spectrum for $T=T_{cold}$ (spectrum in orange color in Figure~\ref{f6}) at $<$1 $\mu$m reproduces the $HST$/ACS measurements (0.6--1.0 $\mu$m) with a spectrum that starts to rise at 0.9 $\mu$m towards shorter wavelengths, which gives an improved fitting quality. 

Regardless of temperature degeneracy, we find that the upper limit of $a$ is consistent, irrespective of the chosen temperature profile. This is because the particle size determines the shape of $Q_{ext}$ across all wavelengths, which is responsible for the slope at $<$1 $\mu$m and an absorption at $>$8 $\mu$m (see Figure~\ref{f3}). Independence of the maximum particle size with respect to temperature tells us that the degeneracy between the two parameters can be broken from the current measurements.  

Secondly, we fix the temperature profile to be $T=T_{dayside}$ and repeat the grid of retrievals for different $\tau$ and $a$, this time shifting the planetary radius by 0.5 per cent from $R_0$. As we discussed in the analysis of temperature sensitivity, it is found that the upper limit of $a$ is still well-constrained even though the radius of the planet varies (Figure~\ref{f5}). Again, the degeneracy between the largest particle size allowed and planetary radius is broken. For a smaller $R_p=0.995R_0$, the optical thicknesses can be larger than the $R_p=R_0$ case that is required to fit the slope in the UV and visible. It is found that the well-fitted solutions also appear at large $\tau$ ($\sim$0.04) and small $a$ ($<$0.01 $\mu$m), the region where a high extinction by aerosols is shown beyond 8 $\mu$m (c.f. spectrum in cyan color in Figure~\ref{f2}). 

We see that an increase in the radius by 0.5 per cent from $R_0$ mimics the effect of an increase in the temperature by 200~K because the main effect from the temperature is an increase in the scale height. A large $R_p$ favors a reduced aerosol $\tau$ at $<$1 $\mu$m to find a solution for the $HST$/STIS and ACS data. In the IR, the continuum absorption by CIA reaches as high as the radius ratio at the bottom of the measured spectra and therefore a small amount of H$_2$O is required to fit the $HST$/NICMOS spectrum. This leads to the shallower best-fit spectrum at wavelengths $>$1 $\mu$m than the best-fit spectrum for the $T=T_{warm}$, $R_p=R_0$ case (see spectra in red and cyan color in Figure~\ref{f6}), even though the $\Delta\chi^2/$N contours are nearly identical. We conclude that constraining the molecular abundance in the terminators with a narrow range of uncertainty is not achievable from the current measurements because the extinction by aerosols is highly degenerate with temperature and planetary radius.

We note here that if the scale height of the model atmosphere is reduced due to a high mean molecular weight found from the retrieval, the model top-of-atmosphere pressure required to reproduce the $HST$/STIS spectrum would be lower than 10$^{-7}$ bar, as the physical extent of the atmosphere is reduced by increasing the mean molecular weight, leading to a poor fit to the entire spectrum. We tested the effect on our results of reducing the model top-of-atmosphere pressure to 10$^{-10}$ bar; we found that, in general, the required total cloud optical depth was reduced by a factor of two in this case, whereas the upper limit on the particle size is unchanged. This implies that although the aerosol solutions are degenerate with the top pressure level, reproducing the slope at wavelengths $<$1 $\mu$m is much more weighted by the particle size as it primarily governs the shape of the spectrum in this wavelength region.

\begin{figure*}
\centering
\includegraphics[width=\columnwidth]{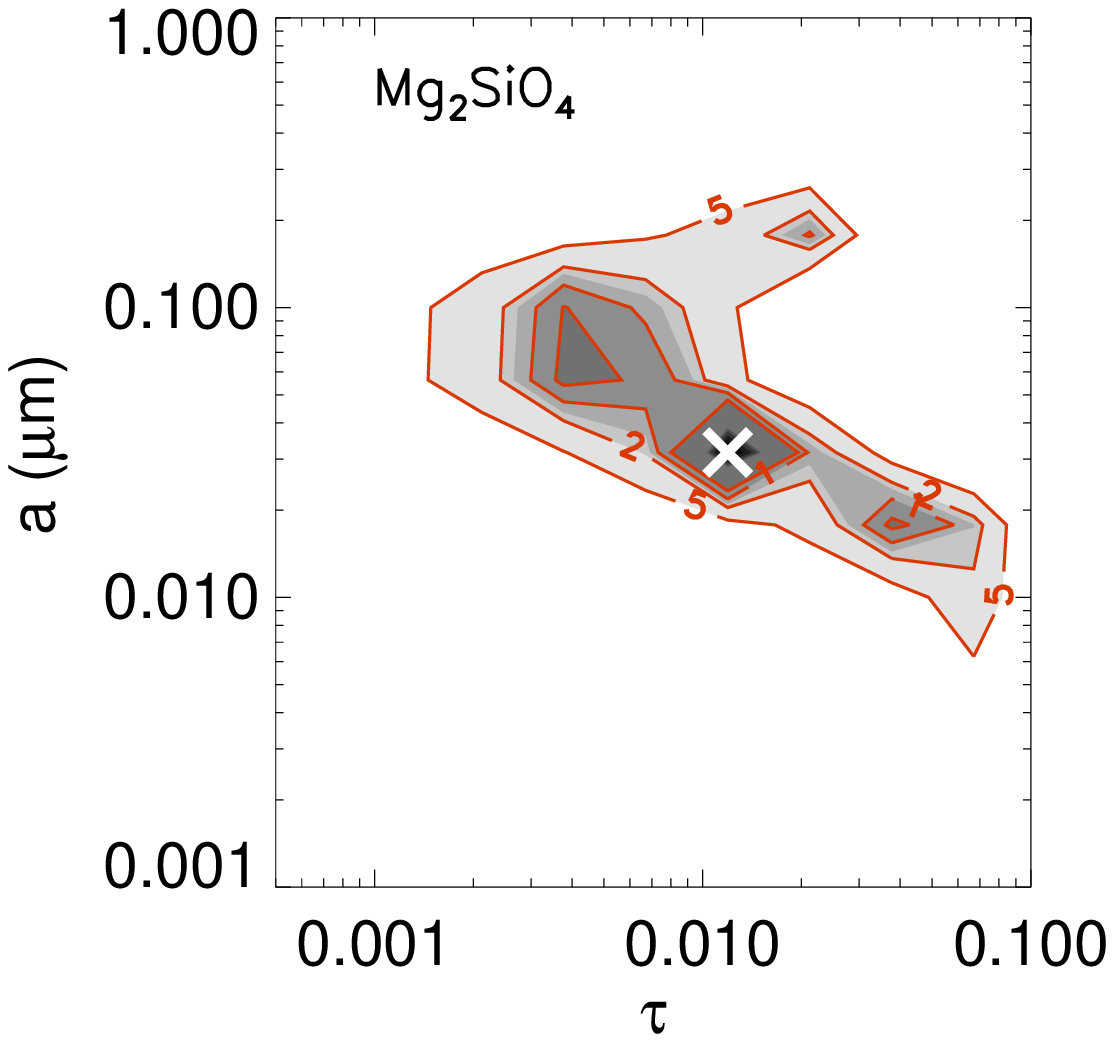}
\includegraphics[width=\columnwidth]{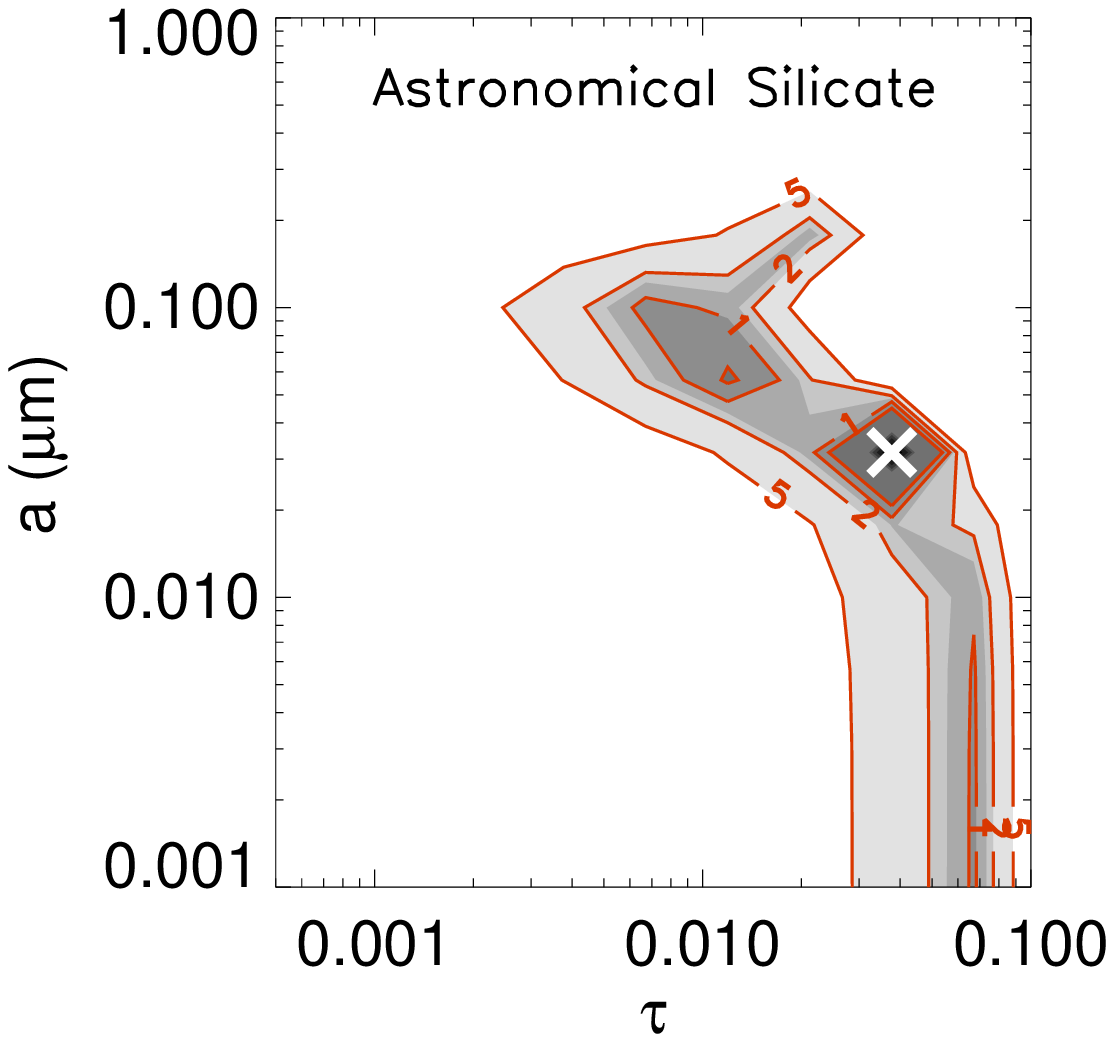}
\includegraphics[width=\columnwidth]{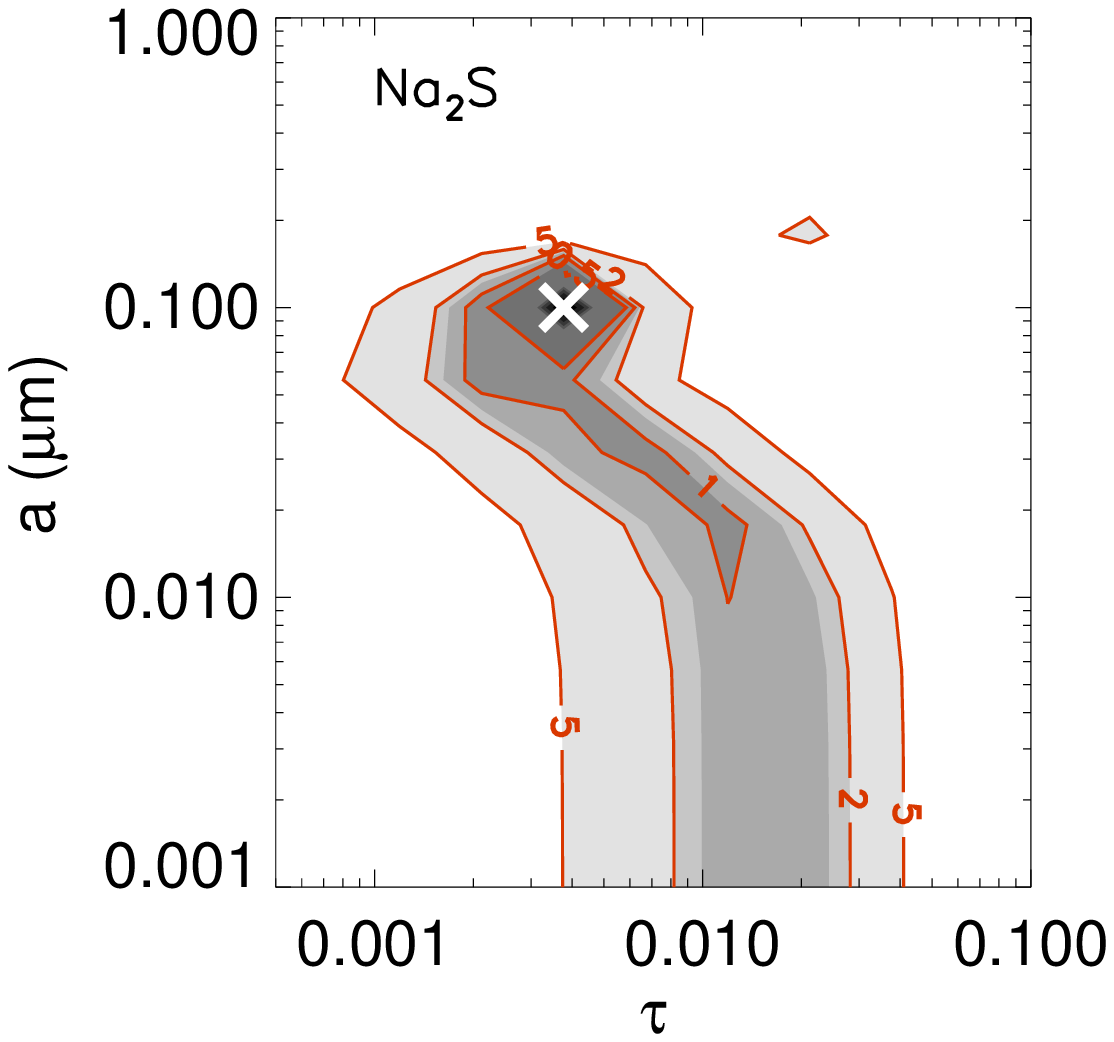}
\includegraphics[width=\columnwidth]{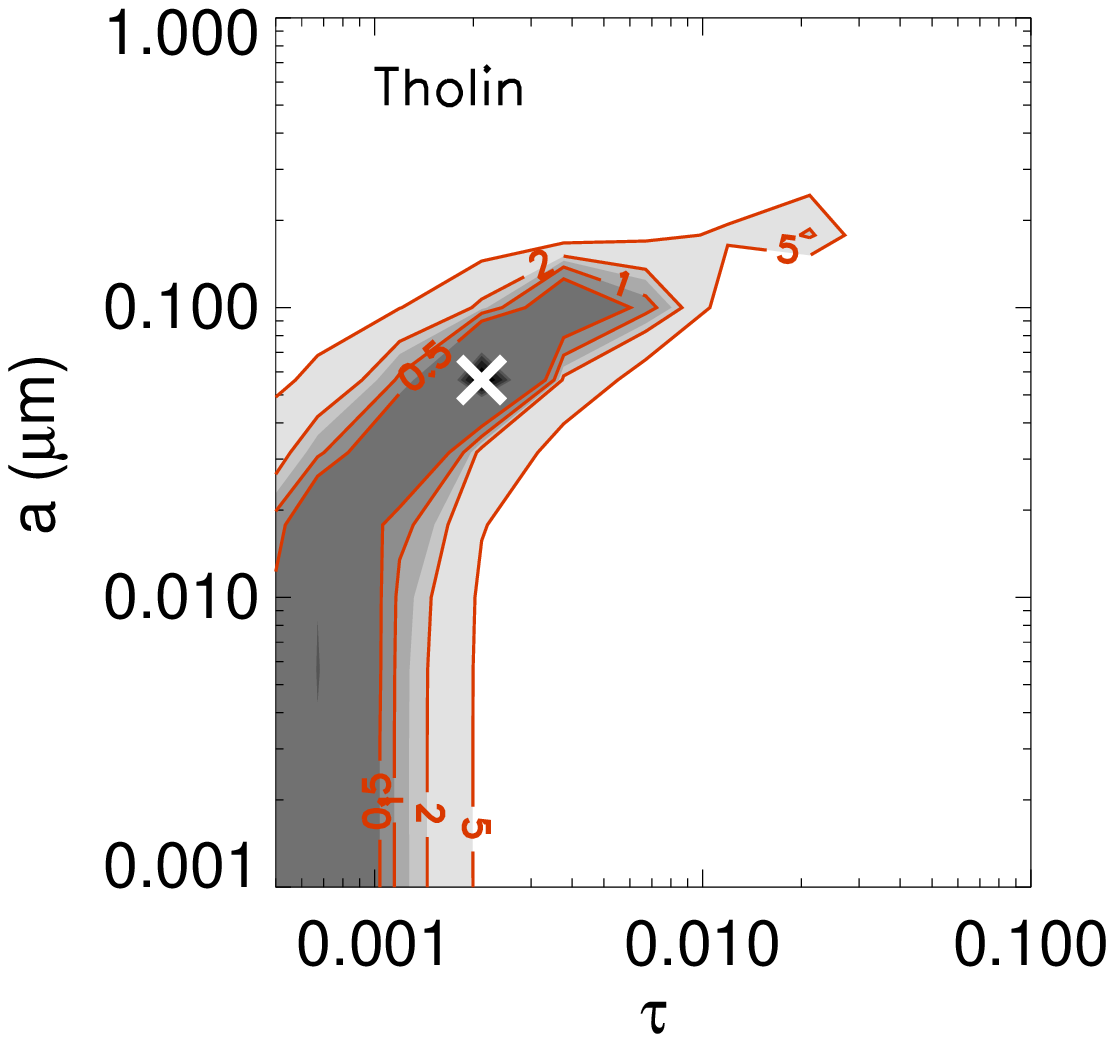}
\caption{The goodness-of-fit, $\Delta\chi^2/$N, contours in aerosol $\tau$ and $a$ space for different candidate materials: Mg$_2$SiO$_4$, astronomical silicate, Na$_2$S, and tholin. } 
\label{f8}
\end{figure*}

\begin{figure}
\centering
\includegraphics[width=\columnwidth]{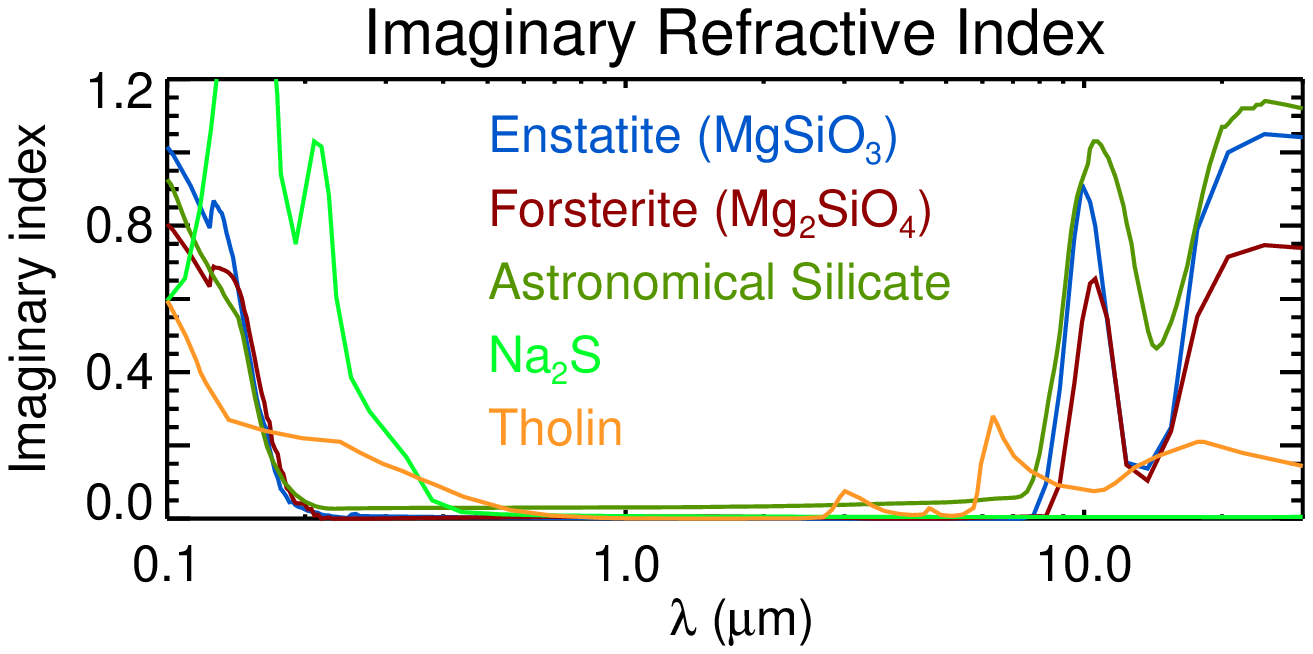}
\includegraphics[width=\columnwidth]{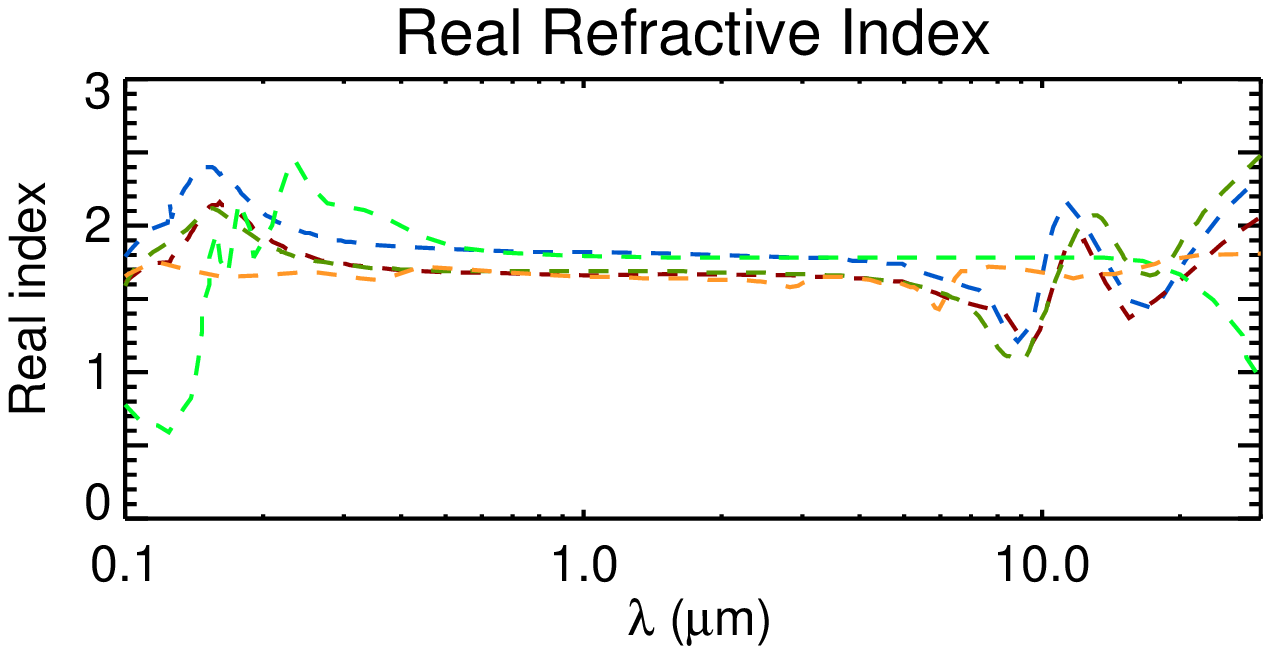}
\includegraphics[width=\columnwidth]{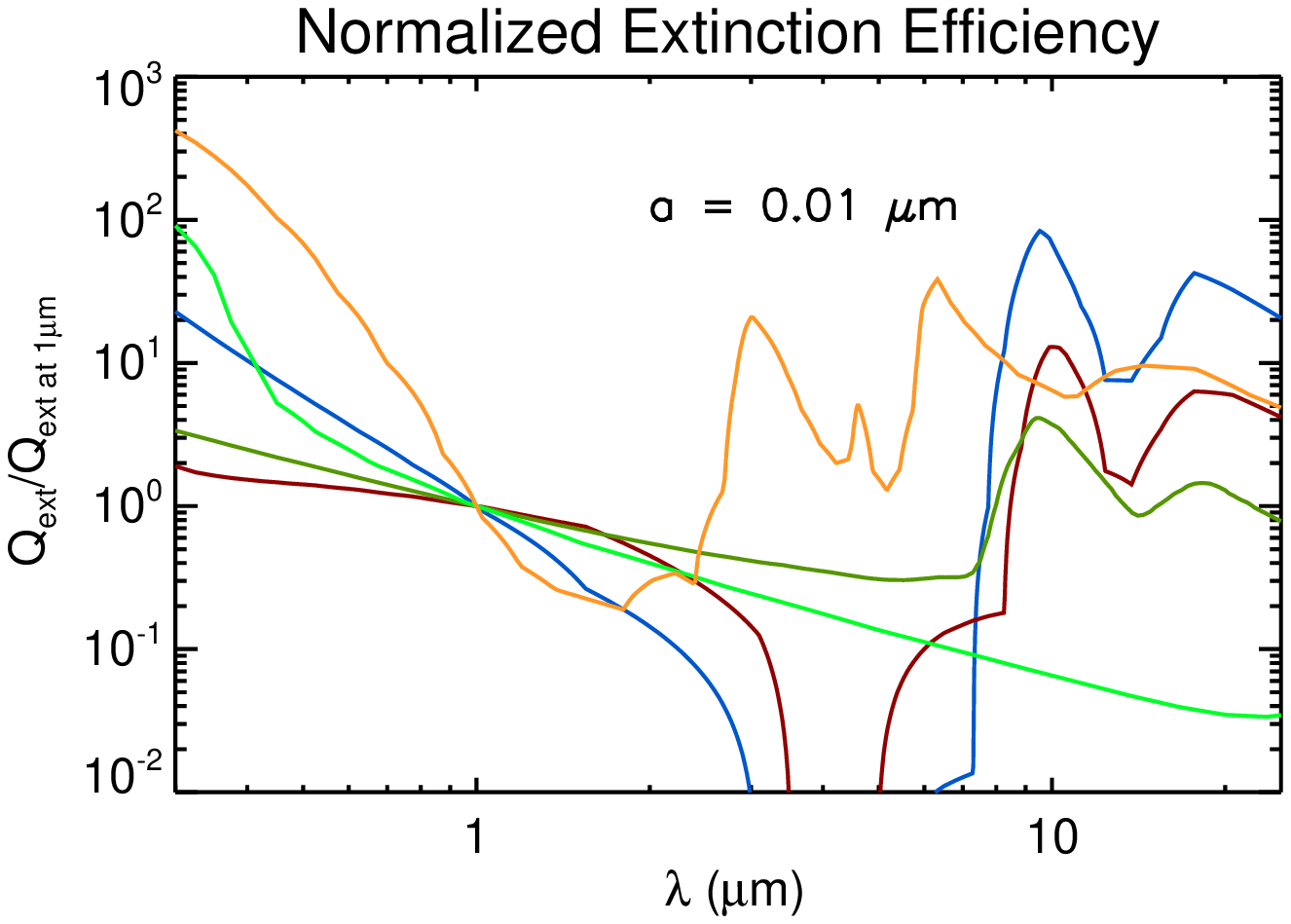}
\caption{Real and imaginary parts of refractive indices for MgSiO$_3$, Mg$_2$SiO$_4$ \citep{sco96}, astronomical silicate \citep{draine1,draine2}, Na$_2$S \citep{mor12} and tholin \citep{kha84}. The bottom panel shows the extinction efficiency ($Q_{ext}$) of the five materials with $a=$0.01 $\mu$m, all normalized to the $Q_{ext}$ at 1 $\mu$m. Tholin has a steep slope at wavelength $<$1 $\mu$m, resulting in a family of solutions with a thin $\tau$ in the contour plot in Figure~\ref{f8} that is different from the other aerosol candidates. }
\label{f9}
\end{figure}

\subsubsection{Aerosol Material}
\label{s322}
In addition to uncertainties associated with the thermal structure and planetary radius, we also consider the differences in aerosol extinction cross section (calculated assuming Mie theory for spherical particles) for different aerosol composition dominating the transmission spectrum of HD 189733b. This constitutes one of the largest uncertainties when modelling the aerosol opacity for a hot Jupiter atmosphere.  We examine how the choice of aerosol materials affect the best-fitting aerosol parameters. Each candidate material has its own complex refractive index spectrum, which determines a wavelength-dependent extinction efficiency ($Q_{ext}$), yielding different $\Delta\chi^2$/N contours in the $\tau$ and $a$ space. Previously, given warm atmospheres under diverse conditions, a range of different candidates have been suggested in the literature. They discussed thermal conditions of condensation of each candidate to constrain the lower deck pressure where a saturation process is triggered \citep[e.g.][]{bur99,ack01}. The aerosol formation mechanism in hot Jupiter atmospheres, however, may differ from that in other warm atmospheres (e.g. low-mass brown dwarfs) due to a strong irradiation from the parent star. We choose aerosols that may be plausible in a warm atmosphere and investigate the effect of $Q_{ext}$ on our aerosol solution. By doing so, we can find the specific shape of $Q_{ext}$, which is required to fit the spectrum best in each wavelength. 

We selected four materials; forsterite (Mg$_2$SiO$_4$), astronomical silicate\footnote{Astronomical silicate is seen ubiquitously in space and it is known that this is a mixture of different silicate grains that might be combined with iron and magnesium.} \citep{draine1,draine2}, sodium sulfide (Na$_2$S), and tholin. Some of these species have a substantially different extinction efficiency compared to the MgSiO$_3$ grains used in our previous retrieval testing. We perform the retrievals for a range of $\tau$ and $a$ with a well-mixed aerosol consisting of each material, for which $Q_{ext}$ is computed from the refractive index as shown in Figure~\ref{f9}. The refractive indices are taken from \citet{sco96} for Mg$_2$SiO$_4$, \citet{draine1} and \citet{draine2} for astronomical silicate, \citet{mon79}, \citet{kha09} and \citet{mor12} for Na$_2$S, and \citet{kha84} for tholin. To study the aerosol degeneracy due to $Q_{ext}$ alone, the vertical temperature profile and the radius of the planet at the bottom of the atmosphere are fixed to be $T=T_{dayside}$ and $R_p=R_{0}$. The $\Delta\chi^2$/N contours for the cases with the four different aerosols are shown in Figure~\ref{f8}. For Mg$_2$SiO$_4$ and astronomical silicate, where the overall shape of $Q_{ext}$ resembles that of MgSiO$_3$ (see Figure~\ref{f9}), the best-fit $\tau$ and $a$ become thicker and smaller than the MgSiO$_3$ case, respectively, showing an improved fit to the spectrum at wavelengths $<$1 $\mu$m (see spectra in orange and cyan color in Figure~\ref{f7}). A small change in the spectrum at short wavelengths leads to a large improvement in goodness-of-fit even though a degraded fitting quality at the $Spitzer$/MIPS channel is found. An increased $\tau$ induces a peak in the spectrum at 10 $\mu$m due to aerosol extinction that produces no significant change in fitting quality because there are no constraints in this wavelength region. The test implies that spectroscopic measurements at $>$ 8$\mu$m contain valuable information to understand the composition of aerosol in a hot Jupiter atmosphere.

On the other hand, we find that the extinction efficiency of Na$_2$S becomes flat at wavelengths $>$1 $\mu$m due to the flat real refractive index and the small imaginary refractive index, whereas a high imaginary index is apparent at wavelengths $<$0.6 $\mu$m. This means that, except for the region at wavelengths $<$0.6 $\mu$m, the best-fit spectrum with Na$_2$S aerosols (spectrum in red in Figure~\ref{f7}) resembles that with the MgSiO$_3$ in all IR regions, where an extinction by MgSiO$_3$ is hardly shown. As a result, only the $HST$/STIS and ACS measurements are useful to characterise the properties of the Na$_2$S aerosols, which adds more uncertainty to $\tau$ and $a$ (see the bottom right panel in Figure~\ref{f8}). This implies that constraining aerosols in an atmosphere is highly dependent on the extinction efficiency of materials and that the properties of aerosols can only really be constrained in wavelength regions where strong extinction features are evident. 

A similar interpretation can be also made in the case of the tholin aerosol, which yields $\tau$ that is thinner than 0.01 in the best-fit case (see the bottom-right panel in Figure~\ref{f8}). The shape of the $\Delta\chi^2$/N contour that is different compared to other candidates also comes from the difference in $Q_{ext}$. For a small particle ($a<$0.1 $\mu$m), the tholin's extinction coefficients at wavelengths shorter than 1 $\mu$m decrease more rapidly with wavelength than the other materials as the bottom panel of Figure~\ref{f9} shows, which induces a high absorption at $<$1$\mu$m, producing good fits to the spectrum even for small $\tau$ ($<$0.001) at 1 $\mu$m. For a large particle ($a\sim$0.1 $\mu$m), tholin becomes also optically active at wavelengths longward of 2 $\mu$m, where the spectra fitted exhibit large absorption features that provide a large deviation from the IR measurements, if $\tau>$0.01. Therefore, we find that the tholin $\tau$ must be very small to reproduce the spectrum, which is constrained by the measurements in between 0.3--24 $\mu$m except $HST$/NICMOS. 

Overall, we found that all of the aerosol materials considered in the present study are capable of fitting the transmission spectrum of HD 189733b. The properties of each aerosol constrained are highly diverse, mainly due to the unique characteristics of $Q_{ext}$ and the divergence of the aerosol properties comes from the uncertainty in condensation state in the terminators. We found that the retrieval of $\tau$ depends on the choice of material, which adds more uncertainty to the retrieval value. Despite the large uncertainty of $\tau$ retrieved, the conclusion on the upper limit of $a$ ($\sim$0.1 $\mu$m) is still robust, regardless of the kind of aerosol material. This is because of the appearance of a peak (the first harmonic) of $Q_{ext}$ at short wavelengths if $a>$ 0.1 $\mu$m, which results in poor fits at the $HST$/STIS bandpasses.

\subsection{Molecular Degeneracy}
\label{s323}
Temperature and other atmospheric parameters such as planetary radius, aerosol properties are highly correlated with gaseous molecules, which are the most significant degeneracy sources in the exoplanet atmospheres. Previously, inverse model studies have shown that the atmospheric molecular abundance in the transiting exoplanets cannot be constrained uniquely from existing transit spectra and therefore the degeneracy between molecules and aerosols, temperature, planetary radius introduces an uncertainty to the atmospheric retrieval method \citep{mad09,lee12,lin10,lin12,lin13,lin14,ben13,bar13}. \citet{bar13} and \citet{ben13} demonstrated that the parameters defining the aerosol deck pressures and its vertical structure of compositional amount modify the molecular abundances, and the uncertain composition of the aerosols increase the uncertainty on the gaseous composition in the retrievals. Furthermore, the choice of value of the planetary radius adds more uncertainty to the determination of molecular abundances. 

\begin{figure}
\centering
\includegraphics[width=\columnwidth]{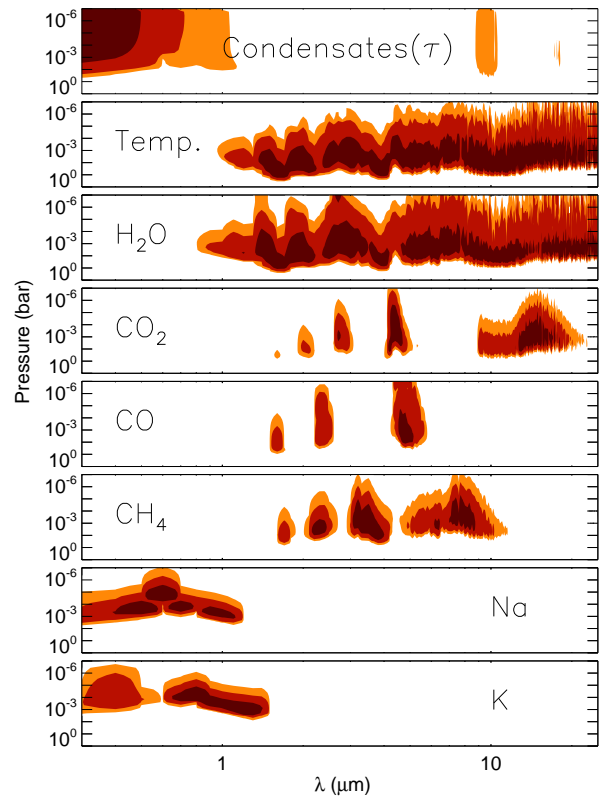}
\caption{Two dimensional spectral sensitivity to aerosol optical depth, temperature, H$_{2}$O, CO$_{2}$, CO, CH$_{4}$, Na and K, plotted against pressure and wavelength. These calculations use the result of the spectrum with the best-fit case in Figure~\ref{f2}.  The contours are normalised to the maximum values in each panel.} 
\label{f10}
\end{figure}

The vertical sensitivity of the spectrum to variations in molecular composition, temperature and aerosol opacity is shown in Figure~\ref{f10}\footnote{This can be defined by taking a partial derivative of spectrum at a wavelength with respect to the parameter of interest, e.g., $\frac{\partial(R_p/R_s)_{\lambda}}{\partial X_i}$.}. It is found that the effect of Na and K are predominant in the UV and visible range, but also influences the near IR and they play a crucial role in fitting the data points at $<$1 $\mu$m. On the other hand, our MgSiO$_3$ aerosol model provides strong sensitivities at short ($<$1 $\mu$m) and long ($>$8 $\mu$m) wavelengths due to a low extinction of Rayleigh slope in the middle range, which is broadly relevant for all other materials tested in the present study. This tells us that the aerosol properties are weakly correlated with the molecules that are spectrally active between 1--8 $\mu$m. In Table~\ref{t1}, we present the ranges of molecular abundances constrained with different aerosol materials. We find that the range of the abundance of H$_2$O for $\Delta\chi^2$/N$<$2 are well constrained with a low uncertainty and this tendency is consistent for all aerosol materials, confirming that there are marginal degeneracies between H$_2$O and aerosol. However, this is only true where the aerosols have negligible influence in the 1--8 $\mu$m range. If the extinction spectrum of aerosols is rather different and there are absorption features in this wavelength range, there would be more uncertainty in the H$_2$O abundance. A possible interpretation of the narrow range of the H$_2$O abundance is that there are only 8 data points in the IR available and they provide strong constraints that determine the molecular abundance. 

For all scenarios that we considered in Table~\ref{t1}, we note that the abundances of CO, CO$_2$, and CH$_4$ have a small range compared to other gaseous molecules. This is because, for any initial atmospheric conditions that are different from the best fit scenario, modifying the abundance of H$_2$O resolves most discrepancies between the measured and fitted spectra in the IR, requiring nearly the same amount of carbon chemicals to fit the data whose opacity contribution to the total spectrum is negligible, i.e., an insignificant contribution to $\chi^2$. We note that retrievals without carbon chemicals marginally change the $\chi^2$/N contours in the $\tau$ and $a$ space and the best-fit fitting quality, e.g., $\chi^2$/N=2.42 to 2.55 for the case of MgSiO$_3$, $T=T_{dayside}$, and $R_p=R_0$. Therefore the existence of CO, CO$_2$, CH$_4$ is highly uncertain. Particularly, the spectral sensitivities to the carbon-bearing molecules are correlated with the signatures of H$_2$O, such that the transmission spectrum is unable to place strong constraints on the presence of CO, CO$_2$ and CH$_4$ due to the degeneracy between them.  Even when the temperature profile is fixed to that derived from the dayside emission spectrum, some of these degeneracies still remain.

Conversely, the range of alkali abundances from retrievals as presented in Table~\ref{t1} appears rather large ($X_{Na}=10^{-8}-0.1$, $X_{K}=10^{-9}-0.1$ for $\Delta\chi^2/$N$<$2).  According to Figure~\ref{f10}, the sensitivities of the spectrum to Na and K in the visible and NIR overlap with that of the aerosols. Their abundances are sensitive to both optical depth and particle size of aerosol, resulting in broad uncertainties in their amounts. This clearly shows a significant degeneracy between alkalis and aerosol--the extinction efficiency of aerosols for any particle sizes smaller than $\sim$0.1 $\mu$m produces a Rayleigh slope at $<$1.2 $\mu$m, where the total optical depth of the atmosphere is the sum of an aerosol $\tau$ ($\propto a^6$ due to Rayleigh scattering) and optical depth by alkalis. Therefore, a small change in aerosol properties ($\tau$ and $a$) leads to a larger change in $X_{alkali}$ to fit the slope. This was suggested theoretically for cloudy exoplanets \citep[e.g.][]{mar99} and is shown in the analysis of the reflection spectrum of this planet \citep{heng13,bar14}.  Despite this degeneracy, an additional retrieval test where Na and K were excluded produces a radius ratio slope at $<$1.2 $\mu$m with $\Delta\chi^2/$N$>$5, compared to the best fit case in Section~\ref{s31}, for all combinations of aerosol $\tau$ and $a$ and H$_2$O (not shown). This test adds more credibility to the presence of Na and K to reproduce the transmission spectrum \citep{sin11,hui12}.   

Figure~\ref{f10} informs us that the spectrum is insensitive to temperature shortward of 1 $\mu$m, where a high sensitivity to the presence of aerosols is shown due to the predominance of Rayleigh scattering.  Longward of 1 $\mu$m, the spectrum is sensitive to both temperature and H$_2$O abundance, confirming that an independent retrieval of temperature is not achievable from the primary transits \citep{tin07a,ben12,bar13}. The molecular degeneracy with temperature therefore increases the uncertainty on the chemistry in the atmosphere because temperature is one of the primary driving forces that determine the opacity of the molecules and therefore the shape of the transmission spectrum.  Given $R_p=R_{0}$, the derived abundance of H$_2$O with $T=T_{dayside}$ for $\Delta\chi^2/$N$<$2, $X_{H_2O}=(0.6-3)\times$10$^{-4}$ is an order of magnitude smaller than $T=T_{cold}$, $(1-100)\times$10$^{-4}$ and larger than $T=T_{warm}$, $(0.07-0.3)\times$10$^{-4}$. Similarly, given $T=T_{dayside}$, the range of the H$_2$O abundance with $R_p=0.995R_0$, $X_{H_2O}=(2-200)\times10^{-4}$ are two and four orders of magnitude enhanced compared to the cases with $R_p=R_0$, $(0.6-3)\times$10$^{-4}$ and $R_p=1.005R_0$, $(0.00006-0.07)\times$10$^{-4}$, thus limiting our ability to determine the abundance of H$_2$O in the terminators within a narrow uncertainty range (see Table~\ref{t1}). It is expected that this uncertainty will become broader when we add the ambiguity of the aerosol material to the error budget.

In summary, the spectrum shows a low sensitivity to gaseous constituents (except H$_2$O) and characterising their abundances is challenging because of (i) a high correlation with aerosol (Na and K) or (ii) the uncertainty in their presence in the atmosphere (carbon molecules). Therefore we conclude that we are unable to robustly determine abundances for gaseous constituents from the spectroscopic measurements used in this study, due to a considerable degeneracy between multiple molecules contained in the model atmosphere and other parameters that determine the shape of spectrum.

\begin{table*}
\renewcommand{\arraystretch}{1.5}
\begin{center}
\scriptsize
\caption{Molecular mixing ratios in the day/night terminators of HD 189733${\rm b}$, constrained for the criterion of $\Delta\chi^2$/N$<$2. The molecular abundances include only the uncertainties associated with the degeneracy due to aerosol properties, temperature, and planetary radius. The uncertainty due to an insufficient number of measurement constraints, i.e., under-constrained problem, is not considered. The numbers for the carbon-bearing molecules are not shown here because their presence in the atmosphere is uncertain.}
\label{t1}
\begin{tabular}{cccc}
\hline
\hline
 & H$_{2}$O (10$^{-4}$) & Na (10$^{-4}$) & K (10$^{-4}$) \\
\hline
MgSiO$_3$ & 0.6--3 & 0.03--700 & 0.2--40 \\
Mg$_2$SiO$_4$ & 0.05--20 & 7$\times$10$^{-4}$--600 & 2$\times$10$^{-5}$--60 \\
Astro. Silicate & 0.02--8 & 6$\times$10$^{-4}$--600 & 9$\times$10$^{-6}$--10 \\
Na$_2$S & 0.7--2 & 20--700 & 0.07--20 \\
Tholin & 0.6--20 & 0.001--800 & 0.005--80 \\
Uncertainty due to aerosol composition & 0.02--20 & 6$\times$10$^{-4}$--800 & 9$\times$10$^{-6}$--60 \\
\hline
$T=T_{dayside}$, $R_p=R_0$ & 0.6--3 & 0.03--700 & 0.2--40 \\
$T=T_{cold}$, $R_p=R_0$ & 1--100 & 0.02--2000 & 1--2000 \\
$T=T_{warm}$, $R_p=R_0$ & 0.07--0.3 & 2--200 & 0.1--4 \\
$T=T_{dayside}$, $R_p=0.995R_0$ & 2--200 & 0.02--2000 & 1--1000 \\
$T=T_{dayside}$, $R_p=1.005R_0$ & 6$\times$10$^{-5}$--0.07 & 2--60 & 0.04--1 \\
Uncertainty due to temperature and planetary radius & 6$\times$10$^{-5}$--200 & 0.02--2000 & 0.04--2000 \\
\hline
\end{tabular}
\end{center}
\end{table*}

\section{Summary and Conclusion}
\label{s4}


We have performed an optimal estimation retrieval to reproduce the transmission measurements for HD 189733b, utilising the broad spectral coverage between 0.3--24 $\mu$m. Our retrieval analysis with a simple 2-parameter aerosol model confirms that we can characterise the terminator regions of HD 189733b using vertically uniform aerosol models that have an optical depth at 1 $\mu$m of between 10$^{-4}$--0.1 and particle radii $\lesssim$0.1 $\mu$m. Exploring a suite of aerosol models for a wide wavelength region, we demonstrated that there are considerable correlations between aerosol parameters and temperature, planetary radius, aerosol composition and even gaseous alkali abundances that are not easily broken--if the temperature or planetary radius increases, the opacity required from the aerosols to fit the transit spectrum decreases. At the same time, Na and K compete to fit the data in the visible and NIR. Moreover, we showed that different materials for the aerosol affect the best-fit $\tau$ and $a$ because they have different extinction efficiencies ($Q_{ext}$). Due to the wide variety of processes that could cause aerosols in a hot Jupiter atmosphere, the choice of candidates increases the degeneracy that leads to a high uncertainty in the retrieval of alkalis and aerosol properties. However, the conclusion on the upper boundary of particle size ($a<$0.1 $\mu$m) is robust, regardless of any degeneracies inherent in the spectrum. Combining the constraints from wavelengths at $<$1 $\mu$m and $>$8 $\mu$m does allow us to draw conclusions on the aerosol solutions for required optical depth and particle size by fitting these transmission spectra.  Furthermore, our aerosol models are consistent with a companion retrieval analysis of the dayside spectrum of HD 189733b \citep[see][]{bar14}.

We found that there is a weak correlation between the molecular abundances (H$_2$O, CO, CO$_2$, and CH$_4$) and the aerosol properties. We confirmed that a strong correlation between H$_2$O and temperature and planetary radius exists, causing a broad uncertainty in the H$_2$O abundance. The existence of carbon-bearing molecules is rather uncertain, because a lack of information is suggested from the measurements, where there exists a strong degeneracy driven by a small number of constraints, i.e., the under-constrained problem.  

We emphasize that the strongest constraints on aerosol properties resulted from measurements at the shortest ($<$1 $\mu$m from $HST$/STIS and ACS) and longest ($>$8 $\mu$m from $Spitzer$/IRAC and MIPS) wavelengths, with a broad trough in the middle due to molecular absorptions.   The long-wavelength region is often overlooked in the literature, but from the refractive index of materials, the aerosol extinction at these wavelengths can be used to break the degeneracy at shorter wavelengths.  A highlight of our study is combining the short and long wavelengths to break degeneracies associated with the aerosol properties.  We suggest that measuring data in these complementary spectral windows will render interpretations of IR spectra more robust. Any future space-based characterization mission should consider this. \\

{\it JL and KH acknowledge the support from the Swiss-based MERAC Foundation, the University of Bern and the University of Z\"{u}rich. PGJI acknowledges the support of the United Kingdom Science and Technology Facilities Council. LNF is supported by a Royal Society Research Fellowship. JKB is supported by the John Fell fund by the University Oxford Press. The calculations were performed using the $zBox4$ computing cluster at the University of Z\"{u}rich thanks to support from Doug Potter, Simon Grimm and Joachim Stadel. We are grateful to David Sing, Frederic Pont, Suzanne Aigrain and Neale Gibson for very helpful discussions.}


\label{lastpage}

\end{document}